\documentclass[12pt,preprint]{aastex}
\usepackage[hyphens]{url}

\shorttitle{Gamma Ray Flare from \object{PSR~B1259$-$63/LS~2883}}
\shortauthors{Khangulyan et al.}

\begin{document}

\title{Post-Periastron Gamma Ray Flare  from \object{PSR~B1259$-$63/LS~2883} as a Result of Comptonization of the 
Cold  Pulsar Wind}

\author{ Dmitry Khangulyan\altaffilmark{1},  Felix
  A. Aharonian\altaffilmark{2,3},   Sergey
 V. Bogovalov\altaffilmark{4},  Marc Rib\'o\altaffilmark{5}  \vspace{5mm}}

\affil{$^{1}$Institute of Space and Astronautical Science/JAXA, 3-1-1
  Yoshinodai, Chuo-ku, Sagamihara, Kanagawa 252-5210,
  Japan\email{khangul@astro.isas.jaxa.jp}}
%
\affil{$^{2}$Dublin Institute for Advanced Studies, 31 Fitzwilliam
  Place, Dublin 2, Ireland\email{felix.aharonian@dias.ie}}
\affil{$^{3}$Max-Planck-Institut f\"ur Kernphysik, Saupfercheckweg 1,
  D-69117 Heidelberg, Germany}  
%
\affil{$^{4}$National Research Nuclear University (MEPHI), Kashirskoe
  shosse 31, Moscow, 115409 Russia\email{svbogovalov@mephi.ru}}
\affil{$^{5}$Departament d'Astronomia i Meteorologia, Institut de
  Ci\`ences del Cosmos (ICC), Universitat de Barcelona (IEEC-UB),
  Mart\'{\i} i Franqu\`es 1, E-08028 Barcelona,
  Spain\email{mribo@am.ub.es}}

\begin{abstract}
  We argue that the bright flare of the binary pulsar
  \object{PSR~B1259$-$63/LS2883} detected by the {\it Fermi} Large
  Area Telescope (LAT), is due to the inverse Compton (IC) scattering of the
  unshocked electron-positron pulsar wind with a Lorentz factor
  $\Gamma_0 \approx 10^4$.  The combination of two effects both linked
  to the circumstellar disk (CD), is a key element in the proposed model.
  The first effect is related to the impact of the surrounding medium
  on the termination of the pulsar wind.  Inside the disk, the
  ``early'' termination of the wind results in suppression of its
  gamma-ray luminosity.  When the pulsar escapes the disk, the
  conditions for termination of the wind undergo significant
  changes. This would lead to a dramatic increase of the pulsar wind
  zone, and thus to the proportional increase of the gamma-ray flux.
  On the other hand, if the parts of the CD disturbed by the
  pulsar can supply infrared photons of density high enough for
  efficient Comptonization of the wind, almost the
  entire kinetic energy of the pulsar wind would be converted to
  radiation, thus the gamma-ray luminosity of the wind could approach
  to the level of the pulsar's spin-down luminosity as reported by the
  {\it Fermi} collaboration.

\end{abstract}



\section{Introduction} \label{sec:intro} 

The pulsar magnetospheres are effective gamma-ray emitters
\citep{fermi_pulsars}. Pulsars can produce potentially detectable
gamma-ray emission also due to the bulk Comptonization of their cold
ultrarelativistic winds. While in the case of isolated pulsars this
radiation component is generally weak, unless the wind is accelerated
(relatively) close to the pulsar \citep{bogovalov00,aharonian12}, in
binary systems the radiation is significantly enhanced due to the
presence of dense target photon field supplied by the optical
companion \citep{ball00,ball01,khangulyan07,khangulyan11}.  Gamma-rays
are produced also after termination of the wind due to the IC
scattering of relativistic electrons accelerated by the termination
shock.  In isolated pulsars the termination of the wind leads to the
formation of the so-called Pulsar Wind Nebulae with a distinct
extended synchrotron and IC emission which can be easily separated
from the radiation of the unshocked wind. In binary pulsars, the
separation of the radiation components produced before and after the
termination of the pulsar wind is a more difficult task; it requires
careful analysis of the spectral and temporal features of two
radiation components.

The most promising object for exploration of processes of formation, acceleration and termination of pulsar winds in binary systems is \object{PSR~B1259$-$63/LS2883}.  It contains a 47.7~ms pulsar orbiting a luminous star in a very eccentric orbit  \citep[eccentricity $e=0.87$, period $P_{{\rm orb}}= 1237$~d, semi-major axis $a_{\rm   2}=7.2\rm \, AU$, see][and references therein]{johnston92,negueruela10}.  The variable TeV gamma-ray emission detected from this object by the {\rm H.E.S.S.}  collaboration \citep{aharonian05,aharonian09}  was in general agreement  with the  predictions  \citep{kirk99}, but
the TeV gamma-ray lightcurve appeared to be  different from expectations.  
Several possible scenarios suggested for interpretation of the reported TeV  lightcurve \citep{khangulyan07},  
predict essentially different gamma-ray fluxes at  MeV/GeV energies, therefore the measurements in this energy band should significantly contribute to the understanding of the hydrodynamics and the particle acceleration processes in binary pulsars.

The recent observations of \object{PSR~B1259$-$63/LS2883} with {\it   Fermi} LAT revealed a complex behavior of the system in GeV gamma-rays \citep{abdo11,tam11}. Close to the periastron passage, the source has been detected with a modest  energy flux of $6\times10^{-11}\rm erg\,cm^{-2}\,s^{-1}$. However, $+30$ days after periastron passage (with largest recorded flux level at $+35$ days),  a spectacular flare has been recorded.  It lasted approximately two weeks with an enhanced flux above 100 MeV at the level of $3\times10^{-10}\rm erg\,cm^{-2}\,s^{-1}$ \citep{abdo11,tam11}.  The flare is  characterized by a sharp increase and a rather smooth decay over approximately $2$ weeks. Importantly, the strong increase of gamma ray flux was not accompanied by a noticeable  change of the 
X-ray flux \citep{abdo11}.
While the  weak GeV flux reported during the periastron passage can be explained by IC scattering of electrons accelerated at the termination shock  and/or  by the bulk Comptonization of the unshocked electron-positron wind \citep{khangulyan11}, the post-periastron flare was a real surprise.  In general,  this flare  represents a unique case in astrophysics when the 
available energy of an object is fully  converted to nonthermal high energy radiation. On the other hand, given the current belief  that the rotational energy of the pulsar is converted to 
kinetic energy of cold ultrarelativistic electron-positron wind, the binary pulsars can in principle operate as 
perfect `100 \%' efficient gamma-ray emitters if the density of the surrounding target photon
field would be high enough  for realization of Comptonization of the pulsar wind in the saturation regime. 
Yet, any explanation of this flare should address several important issues, namely (i) the extraordinary high luminosity of gamma radiation at the level close to the pulsar's spin-down luminosity (SDL), $\dot E=8 \times 10^{35} \ \rm erg/s$; (ii)  lack of  enhanced non-thermal X-ray flux;  (iii) orbital phase of the flare.

\section{The scenario  \label{sec:scenario}}

The very fact of absence of simultaneous activity in the X-ray energy
band suggests that the gamma-ray emission is produced by the pulsar
wind. Indeed, since in the cold wind the electrons move together with
magnetic field, they lose their energy only through the IC scattering,
thus IC gamma-rays are emitted without accompanying synchrotron
radiation. Moreover, for a narrow, e.g. Planckian type distribution of
ambient photons with temperature $T_{\rm r}$, the upscattered
gamma-ray emission will be concentrated within a narrow energy
interval with a characteristic energy $E_\gamma \approx 3 k T_{\rm r}
\Gamma_{\rm 0}^2 \approx 300 (kT_{\rm r}/1 \ \rm eV)\Gamma_{\rm 0,4}^2
\ \rm MeV$ in the Thomson regime ($k T_{\rm r} \Gamma_{\rm 0}/m_{\rm
  e}^2 c^4 \ll 1$) or $E_\gamma \approx m_{\rm e} c^2 \Gamma_{\rm 0}
\approx 5 \Gamma_{\rm 0,4} \ \rm GeV$ in the Klein-Nishina regime
($kT_{\rm r} \Gamma_{\rm 0}/m_{\rm e}^2 c^4 \geq 1$). Here
$\Gamma_{\rm 0}=10^4\Gamma_{\rm 0,4}$ is wind bulk Lorentz factor.

\begin{figure}
\resizebox{\columnwidth}{!}{
\includegraphics[clip]{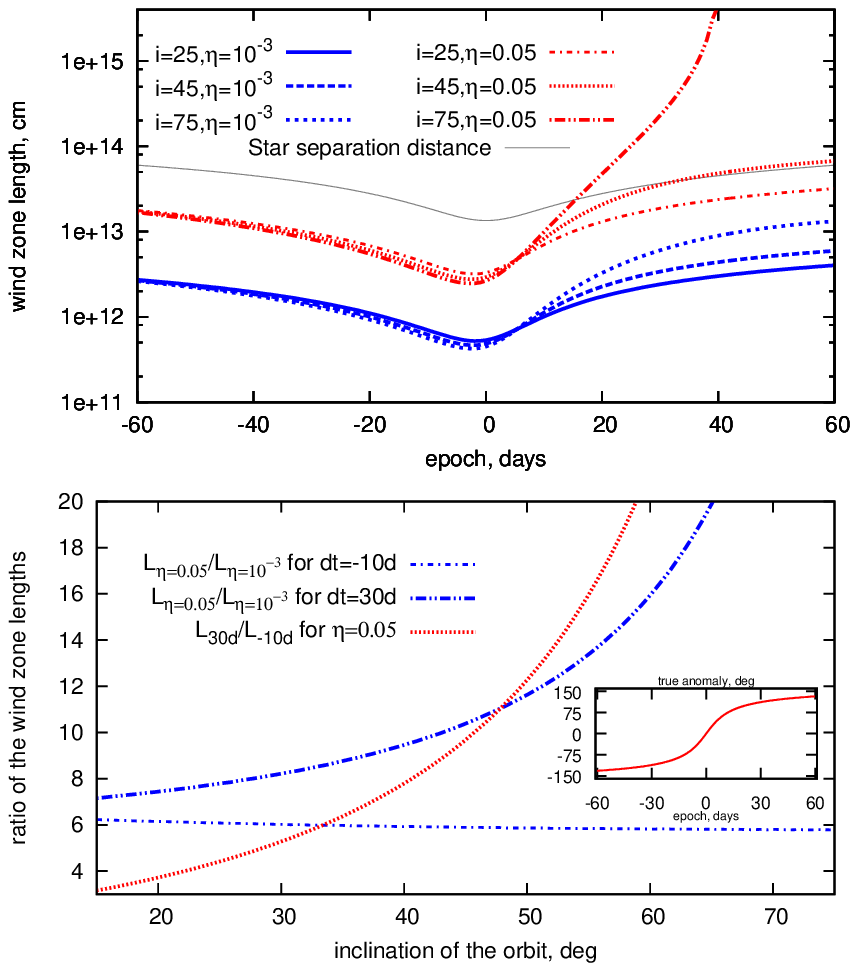}}
\caption{ Upper panel: PWZ lengths  towards the observer as a function of days to periastron. The calculations are
  performed for two values of the $\eta$-parameter: $\eta=10^{-3}$ (CD)
   and $\eta=0.05$ (SW) and  three different
  orbital inclinations: $i=25^\circ$,
  $i=45^\circ$, and $i=75^\circ$.  Bottom panel: ratio of the PWZ lengths 
  for the interaction with SW and CD for
  $-10$~d (dot-dashed line) and $+30$~d (dash-dot-dotted line) to
  periastron passage; and expected ratio of the post- to pre-periastron flares (dotted line). 
  The inset plot shows dependence of the true anomaly (in degrees) on the epoch (in days) to periastron.  }
\label{fig:all}
\end{figure}

The strength of the gamma-ray signal produced by the wind depends on
three parameters: (i) the bulk Lorentz factor of the wind, (ii) the
density of the target photon  field, (iii)  the pulsar wind zone (PWZ)  length  towards the
observer \citep[see e.g.][]{ball01,khangulyan07,sier_bedn08,cerutti08,khangulyan11}.  At least one  of these parameters should experience sudden
changes in order to provide a sharp increase of the gamma-ray signal
as detected by {\it Fermi} LAT.  \citet{bogovalov08} have shown that
the interaction of the pulsar wind can proceed in two different
regimes depending on the ratio $\eta$ of ram pressures of the
interacting pulsar and stellar winds. For $\eta<\eta_{\rm cr}$, where
$\eta_{\rm cr} \simeq10^{-2}$, the pulsar wind termination shock has a
closed structure, and for $\eta>\eta_{\rm cr}$ the termination shock
is expected to be unclosed \citep[on a scale of the binary system, see][]{brb11,brb12}. This
hydrodynamical instability should lead to a fast (on timescales of
$10^3-10^4\rm s$) transformation of the termination
shock. Importantly, in the case of interaction of the pulsar with the
CD $\eta_{\rm d} \sim 10^{-3} \ll \eta_{\rm cr}$,
while at the interaction of the pulsar wind with the stellar wind (SW)
$\eta_{\rm w} \sim 5\times10^{-2}\gg\eta_{\rm cr}$
\citep{khangulyan11}.  Thus,  the pulsar's entrance  to and exit  from
the CD should  lead to fast transformations of the
termination shock.  In Figure~\ref{fig:all}, we show PWZ lengths calculated 
for different orbital inclinations and the $\eta$ parameter \citep[for details see][]{khangulyan11}.

The increase of the PWZ length by a factor of $\sim10$  after its escape of the disk 
naturally explains the  detected dramatic increase of the gamma-ray flux during the flare (see Fig.~\ref{fig:diks}).  
What concerns the  absolute gamma-ray flux with corresponding luminosity comparable to 
the pulsar's SDL,  it requires that the Comptonization of the  wind electrons proceeds in the 
optically thick regime,  i.e. the IC cooling length $l_{\rm IC}$ should not exceed the 
PWZ length of about   $l_{\rm IC} \sim10^{13}$~cm  outside the disk.  In the Thomson regime,
\begin{equation}\label{eq:length}
l_{\rm IC}\simeq10^{14}\Gamma_{\rm 0,4}^{-1} (w_{\rm ph} / \rm erg\,cm^{-3})^{-1} \rm cm\,,
\end{equation}
where $w_{\rm ph}$ is target photons energy density.  Thus, the required energy density of the target photon field is 
\begin{equation}\label{eq:density}
w_{\rm ph}\simeq 10 \Gamma_{\rm 0,4}^{-1} l_{\rm IC, 13}^{-1} \ \rm erg\,cm^{-3}\, ,
\end{equation}
where $l_{\rm IC,13}=l_{\rm IC}/(10^{13}\rm cm)$. 
Although the luminosity of the optical companion LS~2883 is rather
high \citep{negueruela10}, it is still not sufficient for realization
of the optically thick IC scenario \citep{khangulyan11}.  The CD is another source of target photons for Compton scattering of
electrons \citep{soelen11}, which contribution can be significant
(especially in the close vicinity of the disk).  Equations
(\ref{eq:length}) and (\ref{eq:density}) give an estimate of the
required luminosity of the target photon field
\begin{equation}\label{eq:lum}
L_{\rm  ph}> 10^{38}\Gamma_{\rm 0,4}^{-1} l_{\rm IC,13} \rm erg\,s^{-1}\,.
\end{equation}
This is a quite tough requirement, as large as 40 \% of 
the luminosity of the optical companion \citep{negueruela10}. 
One may speculate that such a high photon field is a result of   thermal radiation of 
the heated parts of the CD triggered  e.g.  by
the interaction of  the pulsar with  the disk.  Whether such an effective heating of the disk is possible is 
an open issue.  While the answer to this question  requires detailed hydrodynamical studies,  
below  we simply postulate the existence of such a high density radiation field. 

The requirement on the luminosity of the target photon field given by 
Eq.~(\ref{eq:lum}), can be relaxed assuming larger wind bulk
Lorentz factors and/or smaller values of $l_{\rm IC}$. The Lorentz factor 
cannot much exceed $10^4$,  unless the temperature of radiation is significantly 
less than $10^4 \rm K$ (to explain the reported energy spectrum of gamma-rays).  However,
for $T_{\rm r}\ll10^4\rm K$, the required energy density of the photon field
would exceed the Stefan-Boltzmann limit for black-body radiation. 
Regarding the size of the region
filled with photon field, it cannot be smaller than  $>10^{13}\rm cm$, given that 
during the flare (from $+30\rm d$ to $+60\rm d$ from periastron passage) 
the star separation distance changes from $3.7\times10^{13}\rm cm$ to $6\times10^{13}\rm cm$.

In the framework of the proposed model, one should expect a similar
flare when the pulsar exits the disk during the pre-periastron
epoch. However, due to the differences in the termination shock orientation, the intensity of this flare should be different (see Fig.\ref{fig:sketch}). Assuming that the pulsar exits the disk at epochs symmetric in
terms of true anomaly, the pre-periastron exit should occur 10 days
before the periastron passage. For $\eta=5\times10^{-2}$ the ratio of PWZ lengths
corresponding to $+30$~d and $-10$~d epochs exceeds  4 (see Figs.~\ref{fig:all} and \ref{fig:diks}), 
therefore  the expected flux level is consistent with the {\it Fermi} LAT
upper limits obtained for that period.

Below we present and discuss the results of numerical calculations
performed within the suggested scenario.  The formalism applied to
\object{PSR~B1259$-$63/LS2883} is described in our previous papers
\citep{khangulyan07,khangulyan11}.  The calculations rely on
hydrodynamical simulations of interaction of the pulsar and CD \citep{bogovalov08,bogovalov12}. In this paper we include the effects
related to the presence of the CD which have not been taken
into account in previous studies.

\section{Radiation of  the unshocked pulsar wind}\label{sec:main}

In Figure~\ref{fig:av} we show the spectral energy distribution (SED)
of IC radiation of the pulsar wind averaged over the period of
{\it Fermi} LAT observations  \citep{abdo11}.  
The calculations are performed for different
values of the initial Lorentz factor $\Gamma_{\rm 0}$   
and the $\eta$ -parameter.  These  value
correspond to the following  regimes: $\eta=1$ --  upper
limit; $\eta=0.05$ --  interaction with the SW;
$\eta=10^{-3}$ --  interaction with the CD \citep[see][for details]{khangulyan11}.  The
comparison of calculations with the {\it Fermi} data shows that the IC
radiation of the unshocked wind can achieve the level of
reported fluxes.  However, these fluxes are also rather close
to theoretical predictions for  radiation of  the shocked wind
\citep[see  Figs.~7 and 13 in][]{khangulyan07}.  Unfortunately
the marginal gamma-ray signal obtained during the periastron passage
does not allow a quantitative analysis which 
would give preference to the radiation associated with the 
wind before or after its termination.
On the other hand, the detected gamma-ray fluxes can be treated as upper limits for the unshocked 
wind component. This allow us to constrain the wind's Lorentz factor $\Gamma_0$;
as it follows from Figure~\ref{fig:av},   the interval of  $\Gamma_0 \sim (0.1-3)  \times 10^5$
is excluded by the gamma-ray data.
  
The gamma-ray signal from the pulsar wind associated with the target
photons of the companion star has a smooth orbital phase
dependence \citep{khangulyan11}.  In contrast, the CD can introduce quite sharp temporal features.  Presently, the properties of CDs 
in binary pulsars  are a subject of debate  \citep[see e.g.][]{okazaki11}.  
Here we assume that a CD does exist in  
\object{PSR~B1259$-$63/LS2883}, and that it extends to
$3\times10^{13}$~cm from the star, 
as it follows from the  eclipse  of the pulsed
radio emission observed in 2010  between
$-16$ and $+15$~days  around  the periastron  \citep{abdo11}.
The impact of the disk has been taken into account 
 by introducing two effects: (i) the change of the $\eta$-parameter; and 
 (ii) the presence of an additional  target photon field from the disk itself.

\begin{figure}
\resizebox{\columnwidth}{!}{
\includegraphics[clip]{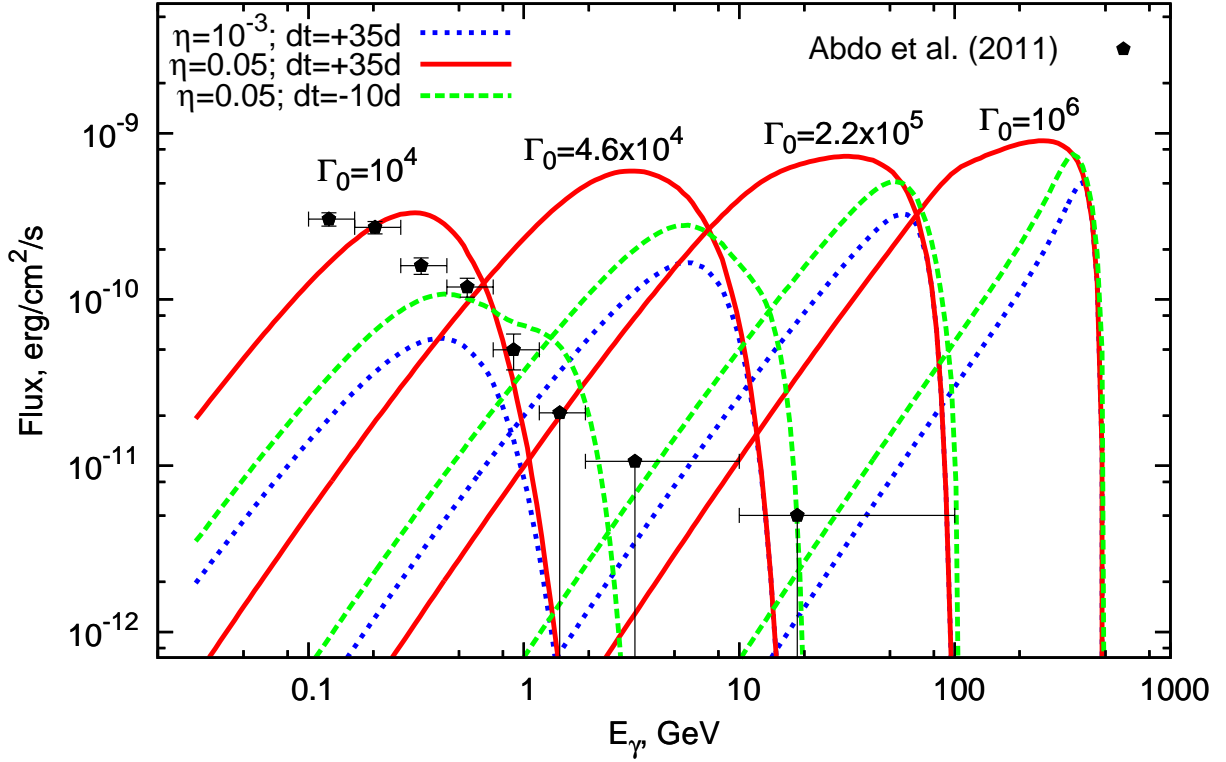}}
\caption{ The spectral energy distribution of the IC radiation of the
  unshocked pulsar wind at the epoch of 35 days after periastron
  passage ($\eta=10^{-3}$--dotted lines; and $\eta=0.05$--solid lines)
  and at the epoch of 10 days before periastron passage ($\eta=0.05$
  -- dashed lines).  The calculations are performed for two target
  photon fields: (i) radiation of the companion optical star with a
  radius $R_*=6.2\times10^{11}\rm cm$ and the surface temperature
  $T_*=3\times10^4\rm K$, and (ii) radiation of the CD,
  which was assumed to be isotropic gray body with temperature $T_{\rm
    r}=10^4\rm K$ and energy density $w_{\rm ph}=2.8\,\rm
  erg\,cm^{-3}$.  Several initial pulsar wind bulk Lorentz factors
  have been assumed: $\Gamma_0=10^4$, $4.6\times10^4$, $2.2\times10^5$
  and $10^6$.  It is assumed that the kinetic energy luminosity of the
  cold ultrarelativistic pulsar wind is equal to the SDL of the
  pulsar. The calculations performed for the model parameters
  $\eta=5\times10^{-2}$ and $\Gamma_0=10^4$ match quite well the
  observational points (pentagons) reported by the {\it Fermi} LAT
   collaboration for the post-periastron flaring
  episode \citep{abdo11}. }
\label{fig:diks}
\end{figure}

\begin{figure}
\resizebox{\columnwidth}{!}{
\includegraphics[clip]{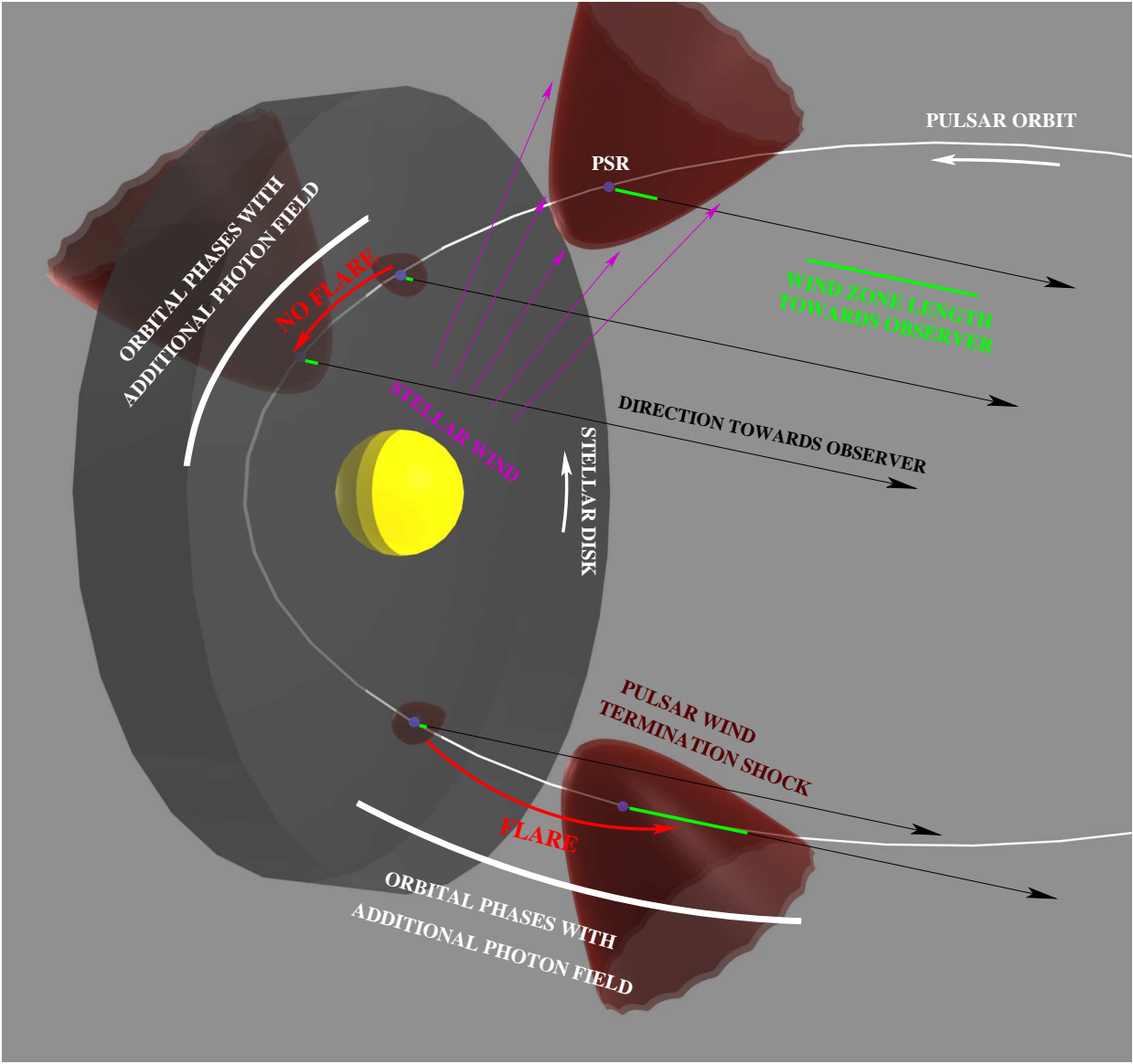}}
\caption{ Sketch of the scenario.}
\label{fig:sketch}
\end{figure}

\begin{figure}
\resizebox{\columnwidth}{!}{
\includegraphics[angle=270,clip]{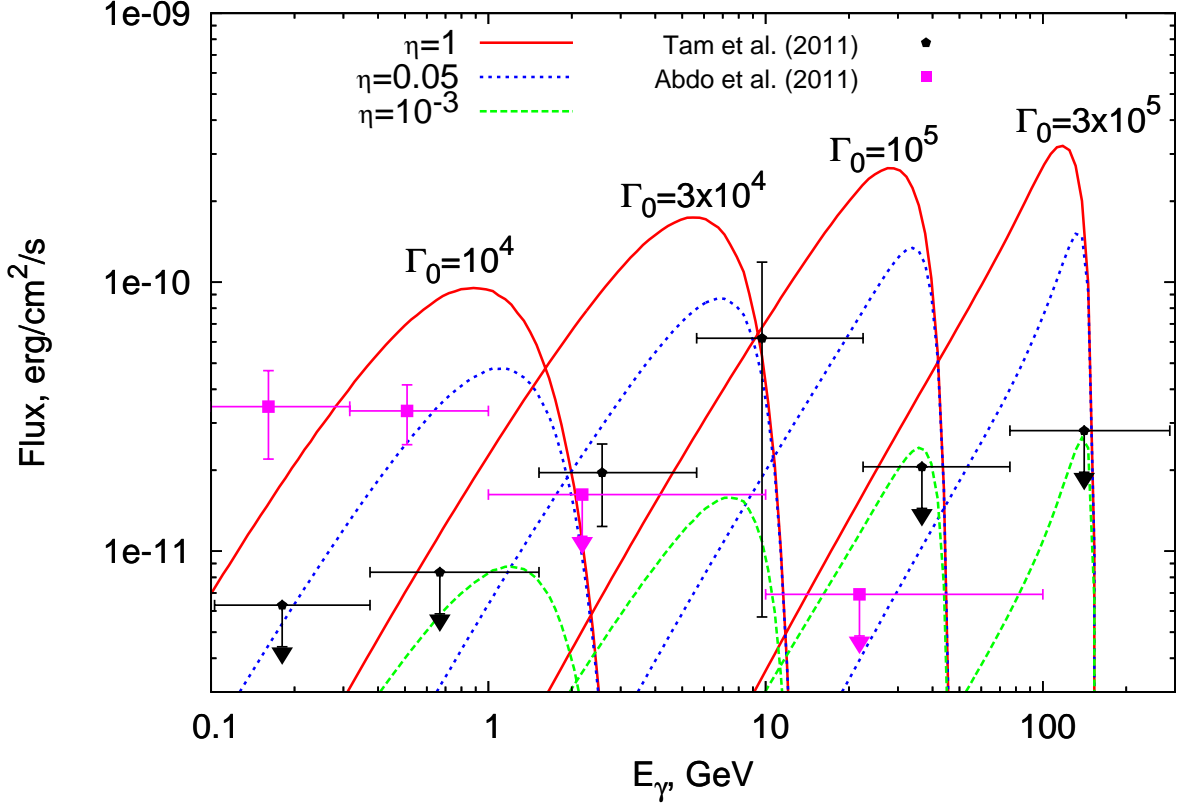}}
\caption{ Spectral energy distributions of IC radiation of the
  unshocked pulsar wind averaged over the  period from $-35$ to $+18$ days around the
  periastron passage. The calculations are  performed for different
  values of the $\eta$-parameter: $\eta=1$ (solid lines), $\eta=0.05$
  (dotted lines, SW) and $\eta=10^{-3}$ (dashed lines, CD), and 
  the pulsar's  wind bulk Lorentz factor: 
  $\Gamma_0=10^4$, $3\times10^4$, $10^5$ and $3\times10^5$. The densities of 
  target photons at different positions of the wind are  calculated assuming  a 
  spherical star with a radius
  $R_*=6.2\times10^{11}\rm cm$ and surface temperature
  $T_*=3\times10^4\rm K$ \citep[see for
  details][]{negueruela10,khangulyan11}. The gamma-ray measurements
  correspond to the  period   of  $[-35, 0]$  days (pentagons) reported by
  \citet{tam11}, and the period of $[-28,18]$ days (squares) reported by \citet{abdo11}.
  }
\label{fig:av}
\end{figure}

 To illustrate the importance of these effects, we performed
 calculations corresponding to the three important cases:(a)
 suppression of the PWZ length by the disk, i.e.
 $\eta=10^{-3}$, $t=+35$ days after periastron passage;
 (b)flare epoch, i.e. $\eta=0.05$, $t=+35$; and (c)
 pre-periastron flare, i.e. $\eta=0.05$, $t=-10$ days before
 periastron passage. The additional target photon field was assumed to
 be gray body with temperature $T_{\rm r}=10^4$~K and energy density
 $w_{\rm ph}=2.8\rm\,erg\,cm^{-3}$. The results of calculations are
 shown in Fig.~\ref{fig:diks}.  The sudden increase of the gamma-ray
 flux by an order of magnitude during the flare is due to the increase
 of PWZ length just after the pulsar leaves the disk. During the first
 1-2 weeks of this epoch the pulsar is located close to the disk, and
 therefore the wind zone is intensively illuminated by the photons of
 the disk.  Apparently, the dominance of the infrared component of the
 disk over the optical radiation of the star, and the upper limit on
 the Lorentz factor of the wind derived from {\it Fermi} LAT
 observations during the periastron passage, explain quite naturally
 the significantly softer spectrum of gamma rays observed during the
 flare \citep{abdo11,tam11}.  From Fig.~\ref{fig:diks} one can see
 that with the assumed set of three model parameters $\eta$, $T_{\rm
   r}$, and $\Gamma_0$, one can explain both the spectral shape and
 the absolute flux of gamma-rays during the flare.  With a slightly
 different choice of parameters than the ones used in
 Fig.~\ref{fig:diks} we can provide, in principle, a perfect fit to
 the observational results. However, given the uncertainties related
 the complex character of the interaction of the wind with the disk,
 the "perfect fit" would be redundant.  Instead, we would like to
 emphasize the most important consequences of this
 interpretation. First, it implies that the Lorentz factor of the wind
 cannot be significantly smaller than $10^4$ given that the energy of
 the IC photons scales as $E_\gamma \propto T_{\rm r} \Gamma_0^2$. On
 the other hand, the {\rm Fermi} LAT observations during the
 periastron passage constrain the wind Lorentz factor to $\Gamma_0
 \leq 10^4$.  Thus, if our interpretation of the flare is correct, the
 wind Lorentz factor should be quite close to $10^4$.

\section{Discussion}

In this letter we propose a model for the recent {\it Fermi}
observations of gamma-rays from \object{PSR~B1259$-$63/LS~2883}. We
assume that gamma-rays are produced at interaction of the cold pulsar
wind with external photon fields. While the Comptonization of the wind
by radiation of the companion star can explain the modest  gamma-ray flux
detected during the periastron passage, the order of magnitude higher
flux of the post-periastron flare requires an additional seed photon
component.  We propose that the radiation of the heated CD can play this role.  The Comptonization of the pulsar wind by the
CD emission proceeds in a quite specific way.  Inside the
disk it is  suppressed  because of the high ram pressure in the disk.
The density of the stellar optical photons
generally dominates over the density of IR photons. But in the
immediate vicinity of the disk, where the IR density can be  still very
high, the optical depth for Compton scattering might  be significantly
larger than 1.  This would lead to a very effective, close to 100\%
efficiency of transformation of the rotational energy of the pulsar to
high energy gamma rays via Comptonization of the cold
ultrarelativistic wind. In the suggested scenario, the departure of
the pulsar wind from the disk will lead to a gradual decrease of the
photon density from the disk and correspondingly to the reduction of
the IC gamma-ray flux on time-scales of days.

During the exit of the pulsar from the disk at the
epoch  before the periastron passage, the termination shock is
expected to expand towards the direction opposite to the observer,
thus we should not expect strong gamma-ray signal. This agrees with 
the {\it Fermi LAT} observations. The main issue for
realization of the proposed scenario is whether the pulsar-disk
interaction can provide the required photon field.  This is a key
question which should be answered by detailed studies of interaction
of the pulsar wind with the CD. This study should also address
another important issue, which is related to the 15 days delay between
the appearance of the pulsed radio emission and the epoch of the GeV
flare. 

If the proposed interpretation is correct,  this would be the second case of 
direct measurement of  a pulsar wind's  Lorentz factor. Recently,  the Lorentz factor 
of order of $10^6$ has been  extracted from the observation of very high energy pulsed 
gamma-ray emission from the Crab Pulsar \citep{aharonian12}. 
Interestingly,  the two orders of magnitude 
smaller Lorentz factor of the pulsar wind  in \object{PSR~B1259$-$63/LS~2883} compared to the 
Crab pulsar, is quite close  the ratio of spin-down luminosities of two pulsars. If this is not just 
a coincidence, the linear dependence $\Gamma_0 \propto L_{\rm SD}$  would imply a universal
production rate of electron-positron pairs of order of $10^{38} \ \rm e^\pm /sec$,   independent of the 
pulsar spin-down luminosity.

Finally, we should note that in principle the gamma-ray flare can be
explained also by the post-shock flow. Since the flux detected by {\it
  Fermi} LAT was close to pulsar's SDL, this
interpretation requires significant Doppler boosting of
radiation. Relativistic flows can be indeed formed at interactions of
the pulsar and SWs \citep{bogovalov08}, and therefore the
broad-band radiation can be affected by the Doppler boosting
\citep{khangulyan08,dubus10}.  The Doppler boosting should amplify
also the X- and TeV gamma ray fluxes.  However, the lack of any
noticeable activity during the flare at other wavelengths
\citep{abdo11} makes this interpretation less likely.  Also, one
should note that the flux level of the Doppler boosted radiation can
easily exceed the limit set by the pulsar's SDL, but
flares with apparent super SDL so far have not been
detected.  On the other hand, the detection of such events in
future observations would rule out the origin of gamma-radiation 
related to the unshocked pulsar wind.

\acknowledgments S.V.B. acknowledges support by Federal Targeted
Program ``The Scientiﬁc and Pedagogical Personnel of the Innovative
Russia'' in 2009-2013 (state contract N 536 on May 17, 2010) and by
grant of Russian Ministry of Science and Education N
2.6310.2011. M.R. acknowledges support by the Spanish Ministerio de
Ciencia e Innovaci\'on (MICINN) under grant FPA2010-22056-C06-02, as
well as financial support from MICINN and European Social Funds
through a \emph{Ram\'on y Cajal} fellowship.

\clearpage



\end{document}